\documentclass[11pt,draft]{article}
 \usepackage{amsfonts}
 \usepackage{amssymb}
 \newfont{\bbbold}{msbm10}

 \def\cE{{\cal E}}

 \def\cL{{\cal L}}

 \newfont{\goth}{eufm10 scaled \magstep1}

 \def\gi{\mbox{\goth i}}

 \def\gn{\mbox{\goth n}}
 
 \def\gp{\mbox{\goth p}}

 \def\gs{\mbox{\goth s}}

 \def\a{\alpha}
 \def\b{\beta}
 \def\c{\gamma}\def\C{\Gamma}
 \def\d{\delta}
 \def\e{\epsilon}
 \def\f{\phi}
 \def\h{\eta}\def\z{\zeta}
 \def\k{\kappa}
 \def\L{\Lambda}
 \def\m{\mu}

 \def\P{\Pi}
 
 \def\s{\sigma}

 \def\x{\xi}
 
 \def\be{\begin{equation}}\def\ee{\end{equation}}
 \def\bea{\begin{eqnarray}}\def\eea{\end{eqnarray}}
 \def\ba{\begin{array}}\def\ea{\end{array}}
 \def\x{\xi}
 
 \def\del{\partial}
 \def\ua{\underline{\alpha}}
 \def\ub{\underline{\phantom{\alpha}}\!\!\!\beta}

 \def\una{\underline a}\def\unA{\underline A}
 \def\unb{\underline b}\def\unB{\underline B}
 \def\unc{\underline c}\def\unC{\underline C}

 \def\unM{\underline M}

 \def\unX{\underline{X}}

 \def\nab{\nabla}

 \def\del{\partial}

 \let\la=\label

 {}

 \def\nn{\nonumber}
 \def\bd{\begin{document}}
 \def\ed{\end{document}}
 \def\bea{\begin{eqnarray}}\def\barr{\begin{array}}\def\earr{\end{array}}
 \def\eea{\end{eqnarray}}
 \def\ft#1#2{{\textstyle{{\scriptstyle #1}\over {\scriptstyle #2}}}}
 \def\fft#1#2{{#1 \over #2}}
 \newcommand{\eq}[1]{(\ref{#1})}
 \def\eqs#1#2{(\ref{#1}-\ref{#2})}
 \def\det{{\rm det\,}}
 \def\tr{{\rm tr}}\def\Tr{{\rm Tr}}

\begin{document}

 \thispagestyle{empty}

 \hfill{KCL-TH-01-44}\\

    \hfill{USITP-01-07}\\

  \hfill{\today}

 \vspace{20pt}

 \begin{center}
 {\Large{\bf Kappa-symmetric higher derivative terms in brane actions}}
 \vspace{30pt}

 {P.S. Howe} \vskip .2cm {Department of Mathematics}\linebreak
 {King's College, London}
  \vskip .5cm
  {and}
  \vskip .5cm
{U. Lindstr\"om}\footnote{e-mail: ul@physto.se} \vskip .2cm{Department of
Physics}, \linebreak
{Stockholm University},
SCFAB\\
{S-106 91 Stockholm, Sweden}\\
\vspace {15pt}

 \vspace{60pt}

 \end{center}

 {\bf Abstract}

Using the superembedding formalism we construct supermembrane
actions with higher derivative terms which can be viewed as
possible  higher order terms in  effective actions. In
particular, we provide the first example of an action for an
extended supersymmetric object with fully $\kappa$-symmetric
extrinsic curvature terms.

 \vfill\leftline{}\vfill \vskip  10pt

 \baselineskip=15pt \pagebreak \setcounter{page}{1}


\section{Introduction.}

Most discussions of $p$-brane effective actions are confined to the
lowest nontrivial order in the momentum expansion. Higher
derivative terms will appear when the expansion is continued,
however, and their supersymmetrisation is of interest either when
considering the possible structure of the effective action or,
taking a different attitude, as possible counterterms/regulators in
a quantum theory of branes/supergravities. 
For $p$-branes the bosonic structure of such terms has been
considered in some cases. In particular, the curvature square terms
of effective D$p$-brane actions that correspond to ${\cal O}(\alpha
')^2$ corrections to string scattering amplitudes were presented in
\cite{Bachas:1999um}, (see also
\cite{Fotopoulos:2001pt},\cite{Wyllard:2001ye},
\cite{Wyllard:2001qe}), and higher $\del F$ corrections to the
open string effective action were given in
\cite{Andreev:1988cb}, (and in, e.g., \cite{Bilal:2001hb}).
Typically, the higher derivative corrections will include
extrinsic curvature contributions. When contemplating the full
theory, this reopens the problem of finding a globally
supersymmetric and locally
$\kappa$-symmetric formulation of a $p$-brane with extrinsic
curvature terms. This is the topic of the present note.

Partly motivated by Polyakov's suggestion that the QCD string
should be viewed as a string with extrinsic curvature terms
\cite{Polyakov:1986cs}, "rigid" strings and $p$-branes were quite
extensively studied in the 80's,
\cite{Kleinert:1986bk,Curtright:1986ed,Lindstrom:1987ps} and early
90's \cite{Polchinski:1992ty}. A $\kappa$-symmetric Green-Schwarz
type "rigid" string was proposed in \cite{Curtright:1987mr}, and a
generalization to higher $p$ was suggested in
\cite{Curtright:1987mh}.  The $\kappa$-symmetry was only shown (for
the string) to second order in the spinorial target space coordinate $\theta$, however, and a fully
$\kappa$-symmetric formulation was only found for the "rigid"
superparticle \cite{Ivanov:1991ub}, \cite{Ivanov:1992cb},
\cite{Gauntlett:1991dw}. (For the 2-brane, spinning, i.e., locally
world-volume supersymmetric, formulations with extrinsic curvature terms
were found
\cite{Lindstrom:1988fr}).

In the $\kappa$-symmetric description of  the "rigid"
superparticle, the $\kappa$-symmetry is embedded in a local
worldline superconformal symmetry, a fact which points to the way
we understand $\kappa$-symmetry today, namely as defined in terms
of the local supersymmetry of the worldsurface in the
superembedding approach to $p$-branes. This latter view is our
starting point in this paper.\footnote{The use of superspace methods to construct higher derivative counterterms in field theory was first used in \cite{Howe:1981th} in the context of $D=4$ supergravity theories}

The superembedding approach yields conditions for super $p$-branes
to be embedded in a $D$-dimensional superspace. Depending on the
values of $d=p+1$ and $D$, the embedding conditions are more or
less restrictive. In the most restrictive cases, the embedded
surface is given on-shell, but for certain other cases, there is
freedom left corresponding to an off-shell world surface multiplet.
We focus on one of these latter cases ($d=3$ in $D=4$), and utilize
the off-shell multiplet to construct our higher order actions. In
constructing these we are to a certain extent guided by the bosonic
higher derivative actions studied previously. The superspace
actions contains an auxiliary scalar superfield, however, and
eliminating this field turns out to be non-trivial and to lead to a
(presumably infinite) series of higher derivative terms in general. A
particularly simple bosonic action results if we choose a specific
superspace action. An interesting feature of this action is that it
only contains the trace-free part of the extrinsic curvatures.

The paper is organized as follows:  in section 2 we give the necessary
background, presenting a brief summary of the superembedding
formalism. In section 3 we specialize to the case of a membrane in $
D=4$ and analyze the torsion equation in detail. Section 4 contains
a discussion of how to construct actions in the embedding context,
and in section 5 this technique is utilized to find higher derivative
actions. Section 6 contains our conclusions and some of our
notation and conventions are collected in an appendix.

\section{Superembeddings}

The superembedding formalism was pioneered in the context of
superparticles in three and four dimensions by Sorokin, Tkach,
Volkov and Zheltukin \cite{stvz}, and has been applied by these and
other authors to various other branes; for a review see
\cite{s}\footnote{For an earlier attempt to use source and target
superspaces, see, e.g., \cite{Gates:1986vk}.}. In
\cite{hs1} it was shown that the formalism can be applied to arbitrary
branes including those with various types of worldvolume gauge fields,
and it was then used to construct the full non-linear equations of
motion of the M-theory 5-brane in an arbitrary supergravity background
\cite{hs2}.

 We consider a superembedding $f:M\rightarrow \unM$, where $M$ is
 the worldvolume of the brane and $\unM$ is the target space.
 Our index conventions are as follows; coordinate
 indices are taken from the middle of the alphabet with capitals
 for all, Latin for bosonic and Greek for fermionic, $M=(m,\m)$,
 tangent space indices are taken in a similar fashion
 from the beginning of the
 alphabet so that $A=(a,\a)$. We denote the coordinates of $M$
 by $X^M=(x^m,\x^{\m})$.
 The distinguished tangent space
 bases are related to coordinate bases by means of the
 supervielbein, $E_M{}^A$, and its inverse $E_A{}^M$.
 We use exactly the same notation
 for the target space but with all of the indices underlined.
 Indices for the normal bundle are denoted by primes, so that
 $A'=(a',\a')$.

 The embedding matrix is the derivative of $f$ referred to the
 preferred tangent frames, thus

  \be
  E_A{}^{\unA}:= E_A{}^M\del_M X^{\unM} E_{\unM}{}^{\unA}
  \la{2.1}
  \ee

 The basic embedding condition is

  \be
  E_{\a}{}^{\una}=0
  \la{2.1.1}
  \ee

Geometrically this states that the odd tangent space of the brane
is a subspace of the odd tangent space of the target superspace at
any point on the brane. To see the content of this constraint we
can consider a linearised embedding in a flat target space in the
static gauge. This gauge is specified by identifying the
coordinates of the brane with a subset of the coordinates of the
worldvolume, so that

 \bea
 x^{\una}&=&\cases {x^a \cr x^{a'}(x,\x)} \\
 \x^{\ua}&=&\cases{\x^{\a}\cr \x^{\a'}(x,\x)}
 \la{2.2}
 \eea

Since

 \be
 E^{\una}=d x^{\una} -{i\over2}d\x^{\ua} (\C^{\una})_{\ua\ub}
 \x^{\ub}
 \la{2.3}
 \ee

in flat space, it is easy to see, to first order in the transverse
fields, that the embedding condition implies that

 \be
 D_{\a} \hat x^{a'}= i(\C^{a'})_{\a\b'} \x^{\b'}
 \la{2.4}
 \ee

where

 \be
 \hat x^{a'}=x^{a'} + {i\over2} \x^{\a} (\C^{a'})_{\a\b'} \x^{\b'}
 \la{2.5}
 \ee

From this equation the nature of the worldvolume multiplet
specified by the embedding condition can be determined. Depending
on the dimensions involved, this multiplet can be one of three
types: (i) on-shell, i.e. the multiplet contains only physical
fields and these fields satisfy equations of motion, (ii)
Lagrangian off-shell, meaning that the multiplet also contains
auxiliary fields (in most cases), and that the equations of motion
of the physical fields are not satisfied, although they can be
derived from a superfield action, and (iii) underconstrained, which
means that further constraints are required to obtain multiplets of
one of the first two types. For thirty-two target space
supersymmetries, the multiplets are always of type (i) or of type
(iii), whereas for sixteen or fewer supersymmetries all three types
of multiplet can occur, with type (iii) arising typically for cases
of low even codimension.

In this paper we shall be concerned with the membrane in $N=1, D=4$
superspace. The worldvolume is therefore $N=1, d=3$ superspace, and
the worldvolume multiplet is an off-shell Lagrangian multiplet
(type (ii)) which is simply a scalar multiplet. This can be seen
from \eq{2.5}. Since the index $a'$ only takes on one value we see
that this equation determines the transverse fermionic superfield
$\x^{\a'}$ to be the fermionic derivative of the transverse bosonic
superfield, but that there are no further constraints on this
field. So the multiplet is a scalar multiplet as claimed. The
components are a scalar, a spin one-half fermion and an auxiliary
scalar.

In order to determine the consequences of the superembedding
condition in the non-linear case, and to find the induced
supergeometry on the brane one uses the torsion equation which is
the pull-back  of the equation defining the target space torsion
two-form. This reads

 \be
 2\nab_{[A} E_{B]}{}^{\unC} + T_{AB}{}^C E_C{}^{\unC} =
 (-1)^{A(B+\unB)}E_B{}^{\unB}
 E_A{}^{\unA} T_{\unA\unB}{}^{\unC}
 \la{2.6}
 \ee

In this equation the covariant derivative acts on both worldvolume
and target space tensor indices. The latter are taken care of by
the pull-back of the target space connection, while the worldvolume
connection can be chosen in a variety of different ways. One also
has to parametrise the embedding matrix. Having done all this, one
can work through the torsion equation starting at dimension zero.
In this way the consequences of the embedding condition \eq{2.1.1}
can be worked out in a systematic and covariant fashion.

The equations of the component or Green-Schwarz formalism can be
obtained by taking the leading ($\x=0$) components of the various
equations that describe the brane multiplet. A key feature of the
formalism is that these component equations are guaranteed to be
$\k$-symmetric because $\k$-symmetry can be identified with the
leading term in a  worldvolume local supersymmetry transformation.
We recall briefly how this works \cite{stvz}. Let $v^M$ be a worldvolume vector
field generating an infinitesimal diffeomorphism. If we write the
superembedding in local coordinates as $f^{\unM}=X^{\unM}(X)$, then
the effect of such a transformation on $\unX (X)$ is

 \be
 \d X^{\unM}= v^M\del_M X^{\unM}
 \la{2.7}
 \ee

If we express this in the preferred bases we find

 \be
 \d X^{\unA}:=\d X^{\unM}E_{\unM}{}^{\unA}=v^A E_A{}^{\unA}
 \la{2.8}
 \ee

Now if we take an odd worldvolume diffeomorphism, $v^a=0$, and use
the embedding condition \eq{2.1.1} we get

 \bea
 \d X^{\una} &=& 0\\
 \d X^{\ua} &=& v^{\a} E_{\a}{}^{\ua}
 \la{2.9}
 \eea

This can be brought to the more usual $\k$-symmetric form if we
define

 \be
 \k^{\ua}:= v^{\a} E_{\a}{}^{\ua}
 \la{2.10}
 \ee

and note that it satisfies

 \be
 \k^{\ua}=\k^{\ub}
 P_{\ub}{}^{\ua}:={1\over2}\k^{\ub}(1+\C)_{\ub}{}^{\ua}
 \la{2.11}
 \ee

where $P$ is the projection operator onto the worldvolume subspace
of the odd tangent space of the target superspace. We can always
write $P=1/2(1+\C)$, and so $\C$ is computable in terms of the
embedding matrix. Substituting this into \eq{2.9} we recover the
normal form of $\k$-symmetry transformations. (Strictly, we should
evaluate this equation at $\x=0$ to get the correct component
form.) The explicit form of $P$ is

 \be
 P_{\ua}{}^{\ub}=(E^{-1})_{\ua}{}^{\c} E_{\c}{}^{\ub}
 \la{2.12}
 \ee

where the inverse is taken in the fermionic tangent space (of
$\unM$).

\section{The membrane in $D=4$}

In this section we shall give the details of the non-linear
embedding of the membrane in $D=4$.

To simplify the discussion we shall take the target space to be
flat. For the embedding matrix we can take, as usual,

\bea E_{\a}{}^{\una}&=&0 \\ E_{\a}{}^{\ua} &=& u_{\a}{}^{\ua} +
h_{\a}{}^{\b'} u_{\b'}{}^{\ua}\ , \label{3.1} \eea

where both $\a$ and $\a'$ are $d=3$ spinor indices taking two
values. We also choose

\be
E_a{}^{\una}=m_a{}^b u_b{}^{\una}\ . \la{3.2} \ee

where

 \be
 {\rm Spin}(1,3)\ni u=\left(\ba{l} u_{\a}{}^{\ua} \\
u_{\a'}{}^{\ua}\ea\right)
 \la{3.3}
 \ee

and where the corresponding element of $SO(1,3)$ is made up by
$u_a{}^{\una}$ and a normal component $u_{3}{}^{\una}$. (Such
variables have been referred to as ''Lorentz harmonics'' in the literature; see, for example, reference
\cite{harmonics}.)

The dimension zero component of the torsion equation \eq{2.6} gives

\be
h_{\a}{}^{\b'}=i\d_{\a}{}^{\b'}h \label{3.4} \ee

and we can choose

\be
T_{\a\b}{}^c=-i(\c^c)_{\a\b}\ , \label{3.5} \ee

if

 \be
 E_a{}^{\una}=f u_a{}^{\una};\qquad f=1+h^2
 \la{3.6}
 \ee

We could have chosen to include the scale factor in the dimension
zero torsion instead, but we make the above choice in order to have
the standard worldvolume dimension zero torsion. In fact, we shall
continue with this policy at higher dimensions so that the torsion
components will be the standard ones of off-shell $N=1, d=3$
supergravity \cite{Brown:1979ma}, that is

 \be
 \ba{cccccc}
 T_{\a\b}{}^{\c} &=& 0 & T_{\a b}{}^{c} &=& 0\\
 &&&&&\\
 T_{ab}{}^c &=& 0 & T_{a\b}{}^{\c}&=&i(\c_a)_{\b}{}^{\c} S
 \ea
 \la{3.7}
 \ee

It is always possible to bring the components of the torsion tensor
to this form by making appropriate choices for the connection and
the even tangent space. The field $S$ will be determined later in
terms of the worldvolume multiplet so that the geometry is indeed
induced.

To analyse the higher-dimensional components of the torsion
equation it is convenient to introduce the $\gs\gp\gi\gn(1,3)$
valued one-form $X=du u^{-1}$. We have, by group theory,

 \bea
 X_{\b}{}^{\c}&=&{1\over4}(\c^{bc})_{\b}{}^{\c}X_{bc} \nn\\
 X_{\b'}{}^{\c'}&=&{1\over4}(\c^{bc})_{\b'}{}^{\c'}X_{bc}\nn\\
 X_{\b}{}^{\c'}&=&-{i\over2}(\c^{b})_{\b}{}^{\c'}X_{b3}\nn\\
 X_{\b'}{}^{\c}&=&-{i\over2}(\c^{b})_{\b'}{}^{\c}X_{b3}
 \la{3.8}
 \eea

where $X_{bc}=-X_{cb}$. Since we have specified the worldvolume
connection already we have to use the torsion equation to solve for
the components of $X$.

It is now a matter of straightforward algebra to go through the
torsion equation and determine the various unknown quantities in
terms of the worldvolume multiplet. If we  write $E_{a}{}^{\ua}$ in
the form

 \be
 E_a{}^{\ua}=\L_a{}^{\a} u_{\a}{}^{\ua} + \psi_{a}{}^{\a'}
 u_{\a'}{}^{\ua}
 \la{3.9}
 \ee

and put

 \bea
 X_{\a,b3} &=& \chi_{\a b3} \nn \\
 X_{\a,bc} &=& \e_{cbd} \chi^d_{\a}
 \la{3.10}
 \eea

we find that all of the dimension one-half fields can be expressed
in terms of $\L_a$ which should be thought of as the derivative of
the transverse fermionic field specifying the embedding. For any
vector-spinor $\z_a$ we put

 \be
 \z_a=\hat \z_a + \c_a \z
 \la{3.11}
 \ee

where the hatted quantity is gamma-traceless. The results are

 \bea
 \hat\psi_a &=& ih \hat \L_a \nn\\
 \hat\chi_a &=& 0\nn\\
 \hat\chi_{a3} &=& - \hat \L_a
 \la{3.12}
 \eea

while for the spin one-half components we have

 \bea
 \psi&=& -{2ih\over 1 + 3h^2}\L \nn\\
 \chi&=& {3h\over 1 +3h^2} \L \nn\\
 \chi_3&=& -{1\over 1+3h^2} \L
 \la{3.13}
 \eea

and

 \be
 \nab_{\a}h=-{3(1+h^2)\over 2(1+3h^2)}\L_{\a}
 \la{3.14}
 \ee

At dimension one we find

 \be
 -i\nab_{\a}\L_{a\b}=(\c^b)_{\a\b}\left(-{1\over2} X_{ab3} -{h\over2}
 X_{ab}+ \h_{ab}h S\right) + \e_{\a\b}\nab_a h
 \la{3.14b}
 \ee

and

 \bea
 X_{a,bc}&:=&\e_{bcd} X_a{}^d;\qquad X_{ab}=-{2h\over f} \e_{abc}
 \nab^c h \nn\\
 X_{a,b3}&=&X_{ab3}=X_{ba3}
 \la{3.15}
 \eea

as well as

 \be
 S=-{h\over 2(1+3h^2)} X_3;\qquad (X_3:=\h^{ab}X_{ab3})
 \la{3.16}
 \ee

In equations \eq{3.14b},\eq{3.15},\eq{3.16} we have omitted fermion
bilinear
terms ($\L^2$ terms) as we shall not need them later, but they can
easily be computed. The leading component of the quantity
$X_{a,b3}$, which is symmetric up to fermion bilinears, can be
thought as the bosonic extrinsic curvature (second fundamental
form) associated with the superembedding.

We note also that

 \be
 -i\nab_{\a}\L_{\b}=\e_{\a\b}({X_3\over6}-hS)-{(1+3h^2)\over
 3f}(\c^a)_{\a\b}\nab_a h
 \la{3.17}
 \ee

again up to fermion bilinears. One can of course also compute the
dimension three-halves torsion and curvature components and the
dimension two curvature in terms of the brane multiplet but we
shall not need these results in the rest of this paper.

\section{Actions}

In \cite{hrs} a method of computing the Green-Schwarz action of any
brane starting form the superembedding formalism was given. It is
similar in some respects to the generalised action principle of
\cite{bsv}. This method works even for on-shell brane multiplets
(with the exception of those that have constrained self-dual tensor
fields), but in cases where the brane multiplet is off-shell one
can find the superspace Lagrangian as well. Some examples of this
in the context of codimension zero superembeddings were given in a
recent paper \cite{Drummond:2001uj}. This sort of construction has
also been discussed in the context of general supersymmetric
theories in \cite{Gates:1997kr}, \cite{ggks}.

We recall briefly the construction of brane actions in the context
of the $D=4$ membrane \cite{hrs,hkss}. There is a closed
worldvolume four-form $W_4$ which in this case is the pull-back of
a target space four-form $G_4$,

 \be
 W_4=f^* G_4 = d( f^* C_3)
 \la{4.1}
 \ee

where $C_3$ is the target space potential three-form. As the
bosonic worldvolume is three-dimensional, and since the de Rham
cohomology of a real supermanifold is the same as that of its body,
it follows that $W_4$ is exact and can be written as $W_4=dK_3$ for
some globally defined worldvolume three-form $K_3$. The three-form
$\cL_3:=K_3-f^*C_3$ is closed, and one can construct the GS action
in terms of $\cL_3$ as follows:

 \be
 S_{GS}=\int_{M_o} dx^m dx^n dx^p \cL_{mnp}(x,0)
 \la{4.2}
 \ee

where $M_o$ is the body of $M$. This action is guaranteed to be
invariant under reparametrisations of $M_o$ and $\k$-symmetry
because these transformations are the leading components of even
and odd worldvolume diffeomorphisms respectively and because
$\cL_3$ is closed.

In fact we can also construct a superfield action from $\cL_3$, or
indeed from any closed three-form $L_3,\ dL_3=0$ \cite{Gates:1997kr},
\cite{ggks}. Since any action will be unaffected by a shift in $L_3$ of
the form
$L_3\rightarrow L_3 + d Y_2$, it follows that we can use this freedom to
choose
$L_{\a\b\c}=0$ and $L_{\a\b c}=i(\c_a)_{\a\b} L$, where $L$ is an
unconstrained scalar superfield. Working through the component
equations of $dL_3=0$ one then finds that

 \bea
 L_{\a\b\c} &=& 0\nn\\
 L_{\a\b c} &=&i(\c_c)_{\a\b} L \nn\\
 L_{ab\c}&=&(\c_{ab})_{\c}{}^{\d} \nab_{\d} L\nn\\
 L_{abc} &=&\e_{abc} L_o
 \la{4.3}
 \eea

where

 \be
 L_o=2i \nab^2L -4S L
 \la{4.4}
 \ee

with

 \be
 \nab^2:=-{1\over2}\nab^{\a}\nab_{\a}
 \la{4.5}
 \ee

so that

 \be
 \nab_{\a}\nab_{\b}L={i\over2}(\c^a)_{\a\b} \nab_a L + \e_{\a\b}
 \nab^2 L
 \la{4.6}
 \ee

We can interpret $L$ as the superfield Lagrangian while $L_o$ is
related to the component Lagrangian which can be obtained from
\eq{4.2}. Equivalently, given a scalar superfield $L$ of the
appropriate dimension, we can construct a closed three-form $L_3$
which has components given by \eq{4.3}. We can then compute the
component Lagrangian using \eq{4.2}.

The Green-Schwarz action for the membrane was discussed from this
point of view in \cite{hrs}. One finds that the only non-vanishing
component of $K_3$ is $K_{abc}:=\e_{abc}K$, where

 \be
 K={1-h^2\over 1+h^2}
 \la{4.7}
 \ee

The contribution this term makes to the component GS action can be
found using

 \be
 K_{mnp}=E_m{}^a E_n{}^b E_{p}{}^c \e_{abc} K
 \la{4.8}
 \ee

so if we define the GS Lagrangian density $\cL(x)$ by

 \be
 \cL(x):=\e^{mnp} \cL_{mnp}(x,0)
 \la{4.9}
 \ee

we find, for the membrane

 \be
 \cL(x)= (\det(E_m{}^a) K )(x,0) + \cL_{WZ}
 \la{4.10}
 \ee

where $\cL_{WZ}$ denotes the Wess-Zumino term which is given by the
pull-back of $C_3$ onto $M_o$.

Now the leading term of $E_m{}^a=e_m{}^a + O(\x)$ is not quite the
usual GS dreibein due to the factor of $f$ included in
$E_a{}^{\una}$, instead we find

 \be
 \bar e_m{}^a = f e_m{}^a
 \la{4.11}
 \ee

where $\bar e_m{}^a$ is the GS dreibein. It is the dreibein for the
GS metric $\bar g_{mn}$ defined by

 \be
 \bar g_{mn} =\cE_m{}^{\una} \cE_n{}^{\unb} \h_{\una\unb}
 \la{4.12}
 \ee

where

 \be
 \cE_m{}^{\una}:=\del_M X^{\unM} E_{\unM}{}^{\una}|
 \la{4.13}
 \ee

the vertical bar denoting evaluation at $\x=0$. (This object is
often called $\P$ in GS notation).

We therefore find

 \be
 \cL(x)=\sqrt{-\det \bar g} f^{-3} K|  + \cL_{WZ}
 \la{4.14}
 \ee

The equation of motion for the auxiliary field $h$ is simply $h=0$,
and substituting this back in \eq{4.14} one finds the usual GS
Lagrangian.

This formula can be adapted for a general Lagrangian defined by a
scalar field $L$ as described above. The bosonic terms in the
component Lagrangian are found from

 \be
 L(x)=\sqrt{-\det \bar g}\left( f^{-3} L_o + \ldots\right)
 \la{4.15}
 \ee

where the dots indicate terms involving contributions involving
fermions.

\section{Higher derivative Lagrangians}

In this section we shall apply the above method to construct higher
derivative Lagrangians for the membrane. The GS Lagrangian is of
standard two-derivative type in the bosonic fields so we shall look
at terms in the bosonic part of the Lagrangian which have four
derivatives. These terms are of extrinsic curvature squared type.
To include them we shall need to introduce a parameter, $\b$ say,
which has dimension $-2$.

The idea is then to write down all possible scalar superfields
constructed from $h$ and $\L_a$ which have the correct dimension,
i.e. 1. There are two possible quadratic terms one can construct
from $\L$,

 \be
 L^{(1)}=-{i\over2}\L^{a\a}\L_{a\a}
 \la{5.1}
 \ee

and

 \be
 L^{(2)}=-{i\over2}(\c^{ab})^{\a\b}\L_{a\a}\L_{b\b}
 \la{5.2}
 \ee

Since $h$ has dimension zero one can multiply each of these by
arbitrary functions of $h$. It is straightforward to calculate the
contributions to $L_o$ that these Lagrangians make. Neglecting
fermion terms we find

 \bea
 -i\nab^{\a}\nab_{\a}L^{(1)}&=& {1\over2}X_{ab3}X^{ab3} +
 {h^2(2+9h^2)\over2(1+3h^2)^2} -{2(1+2h^2 +3h^4)\over f^2}(\nab
 h)^2\la{5.3}\\
 -i\nab^{\a}\nab_{\a}L^{(2)}&=&{1\over2}X_{ab3}X^{ab3}-\left({1\over2}+
 {h^2(2+9h^2)\over(1+3h^2)^2}\right) X_3^2+ {4h^2(2+3h^2)\over
 f^2}(\nab h)^2
 \la{5.3b}
 \eea

A general linear combination of these two terms with $h$-dependent
coefficients will lead to a rather complicated bosonic Lagrangian.
However, there are two particular combinations which generate
simpler results. Clearly we can find functions of $h$ such that
when the two terms multiplied by these functions are added the
terms involving derivatives of $h$ will drop out. For these choices
the $h$ equation of motion would remain algebraic although it would
be very complicated and difficult to solve.  The second possibility
is to take the linear combination $L^{(1)}+{1\over2}L^{(2)}$. In
this case there are still derivative $h$ terms but all the
coefficient functions simplify dramatically. We find

 \be
 L_o^{(1)}+{1\over2}L_o^{(2)}={3\over4}\hat X_{ab3}\hat X^{ab3}-{2\over
 f^2} (\nab h)^2
 \la{5.4}
 \ee

where $\hat X_{ab 3}:=X_{ab3}-1/3\h_{ab} X_3$ is traceless.

To compute the contribution of this term to the bosonic action one
has to change to a coordinate basis and then remember that the
bosonic metric in \eq{5.4} is related by a scale factor to the GS
metric. If one then converts to this metric, and drops all fermion
terms, one finds the following Lagrangian for the sum of the GS and
higher derivative Lagrangian:

 \be
 L(x)=\sqrt{-\det g}\left({1-h^2\over 1+h^2} +{\b\over2}\hat
 K^2-{4\b\over3f^2} (\nab h)^2\right)
 \la{5.5}
 \ee

where

 \be
 K_{mn}{}^3=(\nab_m \del_n x^{\unc} )(u^{-1})_{\unc}{}^3 \nn \\
 \la{5.6}
 \ee

with $\hat K$ being the corresponding traceless tensor. The metric
here is

 \be
 g_{mn}=\del_m x^{\una} \del_n x_{\una}
 \la{5.7}
 \ee

i.e. the standard induced metric for the associated bosonic
embedding. The tensor $\hat K_{mn}{}^3$ is then the traceless
extrinsic curvature for this embedding.

If we now set $h=\tan \phi$  the Lagrangian becomes

 \be
 L(x)=\sqrt{-\det g}( {\b\over2}\hat
 K^2-{4\b\over3} (\nab \f)^2 + \cos 2\f)
 \la{5.5b}
 \ee

Hence the equation of motion for the auxiliary field in this case
is simply the sine-Gordon equation,

 \be
 \nab^2\f -{3\over4\b}\sin 2\f=0
 \la{5.6b}
 \ee

\section{Conclusions}

We have discussed a particular case, the membrane in $N=1, D=4$
flat superspace, where the embedding formalism  allows us to
construct $\kappa$-symmetric extensions with higher derivative
terms rather straightforwardly. Note that the superembedding matrix
$E_A{}^{\unA}$ is invariant under rigid target space supersymmetry
transformations because it is constructed using the invariant
target space supervielbein, and so the invariance of the actions we
have discussed under this symmetry is manifest, provided that we
include the fermion terms, of course. We have also seen that the
component actions we have exhibited can be viewed as superspace
actions which are manifestly invariant under local worldvolume
supersymmetry and this guarantees the invariance of the component
actions under reparametrisations of the bosonic worldvolume and
local $\k$-symmetry. We could, of course, fix the worldvolume gauge
to the physical (static) gauge, say, in which case our actions
would be invariant under non-linearly realised target space
supersymmetry with three-dimensional $N=1$ supersymmetry as the
linearly realised subgroup.

Above we gave an action with extrinsic curvature terms. What about
intrinsic curvatures? In fact, at least at the bosonic level, this
is to a certain extent a matter of how one chooses to express the
action. The Gauss-Codazzi equations relate the intrinsic and
extrinsic curvatures on the world volume to each other and to the
pull-back of the ambient space time curvature. In our case the
ambient (super-) space-time is flat and we may therefore trade
extrinsic and intrinsic curvatures according to
\be
R_{klmn}=K^3_{k[m}K^3_{n]l}~,
\ee
which may be used, e.g., to rewrite the extrinsic curvature term in
(\ref{5.5b}) as
\be
\hat{K}^2=-R+{2\over 3}(K_3)^2~, \ee where $R$ is the curvature
scalar and $K_3$ is the trace of the extrinsic curvature. We can
see from the results of section 3 that a similar situation holds in
superspace. The worldvolume intrinsic curvature superfield, $S$,
has dimension one and could thus appear in the superspace
Lagrangian $L$ multiplied by another function of $h$. However, from
\eq{3.17} we see that $S$ can be expressed as a function $F$, say,
of $h$ times $\nab^{\a} \L_{\a}$ up to fermion bilinears. So we
could integrate such a term by parts to get $F'(h)\nab^{\a}h
\L_{\a}$, whereupon we can use \eq{3.14} to express it as another
function of $h$ times a fermion bilinear. Hence the possible
intrinsic curvature term that could be included in the action does
not give anything different and the actions we have constructed are
the most general at this dimension.

In (\ref{5.5b}) the higher order term is fourth order in
derivatives. In the effective action there will generically be
terms of arbitrary order. The restriction on the order of
derivatives may seem to be built into our choice of quadratic
Lagrangians (\ref{5.1}) and (\ref{5.2}) in Sec. 5. Note however
that a solution to the $h$ field equations for a generic linear
combination of (\ref{5.3}) and (\ref{5.3b}) will generate many
terms of order higher than four when substituted back into the
Lagrangian.

Ideally, when considering higher order terms in the effective
actions, one would like to have access to predictions from string
theory for comparison. In \cite{Bachas:1999um} such results may be
found for $D$-branes to order $\alpha'^2$. This calls for extending
our results to $R^2$ terms and $D$-branes. Quite generally, it
would an interesting task to extend our results to non-flat
backgrounds and, in a first attempt, to those other cases where the
superembedding formalism yields off-shell multiplets, typically
models which have eight or fewer supersymmetries on the brane. \\

Finally, a surprising feature of our Lagrangian (\ref{5.5b}) is the
appearance of the Sine-Gordon equation (\ref{5.6b}). One would like
to know if this is just a coincidence or if, e.g., topological
solutions to this equation have any particular significance.

\section*{Appendix}

In this appendix we gather together some conventions and useful
formulae which were used in the derivation of the equations
presented in the text.

In $D=4$ the metric is $\h_{\una\unb}={\rm diag} (-1,1,1,1)$ and
the $\C$-matrices in a split appropriate to the membrane are

 \bea
 \C^a&=&\c^a\otimes \s_3  \nn\\
 \C^3&=&1\otimes \s_2
 \la{A1}
 \eea

while the charge conjugation matrix $C$ and $\C_5$ are given by

 \bea
 C&=&\e \otimes 1 \nn\\
 \C_5&=& -i(1\otimes \s_1)
 \la{A2}
 \eea

In these equations $\c^a$ denotes the $d=3$ gamma-matrices while
the $\s$-matrices are the Pauli matrices. In $d=3$ the metric is
$\h_{ab}={\rm diag}(-1,1,1)$ and we take $\e^{012}=+1$. The Dirac
matrices obey

 \be
 \c^a \c^b =\e^{abc} \c_c, \qquad (a\ne b).
 \la{A3}
 \ee

The gamma matrices can be chosen to be real, a permissible explicit
choice being $\c^0=i\s_2,\c^1=\s_1,\c^2=\s_3$. The charge
conjugation matrix is $C_3=\e=i\s_2$ and Majorana spinors a real. A
four-dimensional Majorana spinor can be written as

 \be
 \psi=\left(\ba{c} \f\\ \chi\ea\right)
 \la{A4}
 \ee

where the two-component spinors on the right are $d=3$ spinors,
$\f$ being Majorana and $\chi$ being anti-Majorana. (This accounts
for some of the factors of $i$ which appear in the dimension-half
equations of section 3.)

The above formulae for gamma-matrices are valid when the indices
are in the standard position for matrix multiplication, i.e. the
first one down and the second one up. If the second index is
lowered, by means of the four-dimensional charge-cojugation matrix,
one finds

 \be
 (\C^{\una})_{\ua\ub}=\cases{(\C^a)_{\ua\ub}=\left(\ba{cc}
 (\c^a)_{\a\b} & 0 \\ 0& -(\c^a)_{\a'\b'}\ea\right) \cr
 \phantom{---------------------------}\cr
 (\C^3)_{\ua\ub}=\left(\ba{cc} 0 &\ \ -i\e_{\a\b'}\\
 +i\e_{\a'\b} &\ \  \ \ \ \ \ 0\ \ \ \ \ea\right) }
 \la{A5}
 \ee

For the gamma-matrices with two vector indices we have, with the
indices in normal matrix position,

 \be
 \C^{\una\unb}=\cases{
 \C^{ab}=\c^{ab}\otimes 1 \cr
 \C^{a3}=-i\c^a \otimes\s_1}
 \la{A6}
 \ee

With both indices down these become

 \be
 (\C^{\una\unb})_{\ua\ub}=\cases{
 (\C^{ab})_{\ua\ub}=\left(\ba{cc} (\c^{ab})_{\a\b} & \ \ \ \  0\\0&
 \ \ \ \ \ (\c^{ab})_{\a'\b'}\ea\right)\cr
\phantom{---------}\cr
 (\C^{a3})_{\ua\ub}=\left(\ba{cc} 0 &-i(\c^a)_{\a\b'}\\
 -i(\c^a)_{\a'\b} & 0\ea\right)}
 \la{A7}
 \ee

$\C_5$ with indices down is

 \be
 (\C_5)_{\ua\ub}=-i\left(\ba{cc} 0 & \e_{\a\b'}\\
 \e_{\a'\b} & 0\ea\right)
 \la{A8}
 \ee

while the charge conjugation matrix, which naturally has its
indices down, is

 \be
 C_{\ua\ub}=\left(\ba{cc} \e_{\a\b} & 0\\
 0 & \e_{\a'\b'}\ea\right)
 \la{A9}
 \ee

\section*{Acknowledgements}

This article represents joint work carried out under EU contract
HPNR-CT-2000-0122; this work was also supported in part by PPARC
through SPG grant 68 and in part by NFR grant 650-1998368.

\end{document}